\documentclass{article}
\usepackage[pdftex]{graphicx}
\usepackage{float}

\begin{document}

{\Large Stochastic modeling and estimation }

{\Large of COVID-19 population dynamics\bigskip }

\textbf{Nikolay M. Yanev}\bigskip $^{1}$, \textbf{Vessela K. Stoimenova}$%
^{2} $\textbf{, Dimitar V. Atanasov}$^{3}$

\textbf{Abstract.}

The aim of the paper is to describe a model of the development of the
Covid-19 contamination of the population of a country or a region. For this
purpose a special branching process with two types of individuals is
considered. This model is intended to use only the observed daily statistics
to estimate the main parameter of the contamination and to give a prediction
of the mean value of the non-observed population of the contaminated
individuals. This is a serious advantage in comparison with other more
complicated models where the observed official statistics are not
sufficient. In this way the specific development of the Covid-19 epidemics
is considered for different countries.\bigskip

2010 Mathematics Subject Classification: Primary 60J80

Secondary 60F05; 60J85; 62P10

Key words: branching processes, estimation\bigskip

\textbf{1. Introduction.}

Nowadays the Theory of Branching Processes is a powerful tool for modeling
the development of populations whose members have an ability of reproduction
following some stochastic laws. The objects may be of different types and
physical nature. From nuclear reaction and cosmic rays to cell proliferation
and digital information, branching stochastic models are used to explain
very interesting real-world stochastic phenomena. Branching processes have
serious applications in Physics, Chemistry, Biology and Medicine,
Demography, Epidemiology, Economics an so on. The basic models and
analytical results are presented in many books and a lot of papers. We would
like to point out the monographs $[1-5]$ among the others. Some applications
of branching processes in Biology and Medicine are presented in $[4]$ and $%
[6]$. \ Some basic estimation problems are considered in $[7].$

The aim of the present paper is to describe an adequate model of the
development of the Covid-19 contamination in the population. For this
purpose a special branching process with two types of contaminated
individuals is constructed and considered day by day. In this way we are
able to use the observed statistics of the Covid-19 daily registered
contaminated individuals and to estimate the main parameter of
contamination. In fact this parameter $m$\ represents the mean value of the
contaminated individuals by one individual per day. Using the observed
statistics some methods for estimation are proposed and the corresponding
graphics are presented. In this way we are able to give a prediction of the
possible development of the mean value of the contaminated individuals. The
modeling and estimation is performed for Bulgaria and some other countries:
Italy, France, Germany, Spain globally. Of course, the model can be applied
to any other country or region. In the paper the results are presented in
detail for Bulgaria, Italy and globally. Results for other countries,
additional information, reports and plots, related to this research can be
found on http://ir-statistics.net/covid-19.

The theoretical model is described in detail in Section 2. The p.g.f.'s and
the mathematical expectations are obtained. Regardless of its simplicity the
model has a great advantage using only the observed official statistics of
the lab-confirmed cases. The estimation problems are presented in Section 3.
Some conclusive remarks are given in Section 4.

\textbf{2 Two-type branching process as a model of Covid-19 population
dynamics.}

What can be observed? - Only that part of the contaminated individuals who
became ill or who are discovered as a result of medical tests. Every day
only the statistics of this registered part of all contaminated individuals
is available.

To describe this situation we can consider a two type branching process $%
\{Z_{1}(n),Z_{2}(n)\}$ where type $T_{1}$ are contaminated (but still
healthy) individuals who don't know that they are Covid-19 infected and type 
$T_{2}$ of discovered with Covid-19 virus individuals (and this is our real
statistics). Every individual of type $T_{1}$ (contaminated) produces per
day a random number of new individuals of type $T_{1}$ (contaminated) or
only one individual of type $T_{2}$ (more precisely, in this case the
individual type $T_{1}$ is transformed into an individual type $T_{2}).$
Note that $T_{2}$ is a final type, i.e. the individuals of this type don't
take part in the further evolution of the process because they are isolated
under the quarantine.

Let $\mathbf{\xi }_{1}=(\xi _{1}^{(1)},\xi _{2}^{(1)})$ be the offspring
vector of type $T_{1}.$ Then the offspring joint probability generating
function (p.g.f.) of type $T_{1}$ can be defined as follows:\bigskip \newline
$h_{1}(s_{1},s_{2})=\mathbf{E(}s_{1}^{\xi _{1}^{(1)}}s_{2}^{\xi
_{2}^{(1)}})=p_{0}+\sum_{j=1}^{k}p_{j}s_{1}^{j}+qs_{2},q=1-%
\sum_{j=0}^{k}p_{j},h_{1}(1,1)=1,$\bigskip \newline
where $|$ $s_{1}|\leq 1,|s_{2}|\leq 1.$

Obviously $h_{2}(s_{1},s_{2})\equiv 1$ because type $T_{2}$ has $(0,0)$
offspring.

Note that $p_{0}$ is the probability that type $T_{1}$ goes out of the
reproduction process (the individual becomes healthy or goes out of the
country, i.e. emigrates), $p_{j}$ is the probability to produce new $j$
contaminated individuals of type $T_{1}$ and $q$ is the probability that the
individual type $T_{1}$ is confirmed ill (or dead). In other words, $q=%
\mathbf{P}\{T_{1}\rightarrow T_{2}\},$ i.e. with probability $q$ an
individual of type $T_{1}$ is transformed into an individual of type $T_{2}.$
One can obtain also that the marginal p.g.f. are\newline
$\mathbf{E(}s_{1}^{\xi
_{1}^{(1)}})=h_{1}(s_{1},1)=p_{0}+\sum_{j=1}^{k}p_{j}s_{1}^{j}+q=1-%
\sum_{j=1}^{k}p_{j}(1-s_{1}^{j}),$\newline
$\mathbf{E(}s_{2}^{\xi _{2}^{(1)}})=h_{1}(1,s_{2})=1-q+qs_{2}.$

If we assume that $Z_{1}(0)>0$ and $Z_{2}(0)=0$ then for $n=1,2,...$

$Z_{1}(n)=\sum_{j=1}^{Z_{1}(n-1)}\xi _{1}^{(1)}(n;j),$

$Z_{2}(n)=\sum_{j=1}^{Z_{1}(n-1)}\xi _{2}^{(1)}(n;j),$ \newline
where \ the vectors $\{(\xi _{1}^{(1)}(n;j),\xi _{2}^{(1)}(n;j)\}$ are
independent and identically distributed (iid) as $(\xi _{1}^{(1)},\xi
_{2}^{(1)}).$

\textit{Interpretation: }$Z_{1}(n)$ is the total number of individuals (type 
$T_{1}$) in the $n$-th day contaminated by the individuals of the $(n-1)$-th
day; $Z_{2}(n)$ is the total number of the registered with Covid-19
individuals (type $T_{2}$) in the $n$-th day. The process starts with $%
Z_{1}(0)$ contaminated individuals, where $Z_{1}(0)$ can be an
integer-valued random variable with a p.g.f. $h_{0}(s)=Es^{Z_{0}}=%
\sum_{k=1}^{\infty }p_{0k}s^{k},$ $|s|\leq 1$, or $Z_{0}=N$ for some integer
value, $N=1,2,...$ . The random variable $\xi _{1}^{(1)}(n;j)$ is the number
of contaminated individuals (type $T_{1}$) in the $n$-th day infected by the 
$j$-th contaminated individual from the $(n-1)$-th day, $%
j=1,2,...,Z_{1}(n-1) $. Similarly the random variable $\xi _{2}^{(1)}(n;j)$
is the number of the confirmed contaminated individuals (type $T_{2})$ in
the $n$-th day transformed by the $j$-th contaminated individual from the $%
(n-1)$-th day, $j=1,2,...,Z_{1}(n-1)$.

Note that $\mathbf{P}\{\xi _{2}^{(1)}(n;j)=0\}=1-q$ and $\mathbf{P}\{\xi
_{2}^{(1)}(n;j)=1\}=q.$ Hence $Z_{2}(n)\in Bi(Z_{1}(n-1,q),$ i.e. \newline
$\mathbf{P}\{Z_{2}(n)=i|Z_{1}(n-1)=l%
\}=(_{i}^{l})q^{i}(1-q)^{l-i},i=0,1,...,l;l=0,1,2,...$

In other words the probability $q$ can be interpreted as a proportion of the
confirmed individuals in the day $n$ among all contaminated individuals in
the day $n-1$.

Let $h_{0}(s)=\mathbf{E}s^{Z_{1}(0)},$ $F_{1}(n;s)=\mathbf{E(}s^{Z_{1}(n)}),$
$F_{2}(n;s)=\mathbf{E(}s^{Z_{2}(n)}).$\ Introduce the following p.g.f.%
\newline
$h^{\ast }(s)=h_{1}(s,1)=q+p_{0}+\sum_{j=1}^{k}p_{j}s^{j},$ $\widetilde{h}%
(s)=h_{1}(1,s)=1-q+qs.$

Then it is not difficult to check that for $n=0,1,2,...,$ we are able to
obtain the p.g.f. of the process:\newline
$F_{1}(n;s)=\mathbf{E(}s^{Z_{1}(n)})=F_{1}(n-1;h^{\ast
}(s))=F_{1}(0;h_{n}^{\ast }(s))=h_{0}(h^{\ast }(h^{\ast }(...(h^{\ast
}(s))...))),$\newline
$F_{2}(n;s)=\mathbf{E(}s^{Z_{2}(n)})=F_{1}(n-1;\widetilde{h}(s))=F_{1}(0;%
\widetilde{h}_{n}(s))=h_{0}(\widetilde{h}(\widetilde{h}(...(\widetilde{h}%
(s))...))),\newline
$where the p.g.f. $h_{n}^{\ast }(s)$ and\ $\widetilde{h}_{n}(s)$ are
obtained after $n$ compositions of the p.g.f. $h^{\ast }(s)$ and $\widetilde{%
h}(s)$ \newline
$h_{n}^{\ast }(s)=h^{\ast }(h^{\ast }(...(h^{\ast }(s))...)),h_{0}^{\ast
}(s)=s;\widetilde{h}_{n}(s)=\widetilde{h}(\widetilde{h}(...(\widetilde{h}%
(s))...)),\widetilde{h}_{0}(s)=s.$

Let $m=\frac{d}{ds}h^{\ast }(s)|_{s=1}=\mathbf{E}\xi
_{1}^{(1)}=\sum_{j=1}^{k}jp_{j}$ be the mean value of the new contaminated
individuals by one contaminated individual (c.i.). Note that $\frac{d}{ds}%
\widetilde{h}(s)|_{s=1}=\mathbf{E}\xi _{2}^{(1)}=q$ is the mean value of the
registered contaminated individuals by one c.i. Introduce also $m_{0}=%
\mathbf{E}Z_{1}(0)=\frac{d}{ds}h_{0}(s)|_{s=1}.$ Therefore

$M_{1}(n)=\mathbf{E}Z_{1}(n)=m_{0}m^{n},n=0,1,2,...,$

$M_{2}(n)=\mathbf{E}Z_{2}(n)=q\mathbf{E}Z_{1}(n-1)=qm_{0}m^{n-1},n=1,2,...;%
\mathbf{E}Z_{2}(0)=0.$

Note that we can observe only $Z_{2}(1),Z_{2}(2),...,Z_{2}(n).$

What can be estimated with these observations?

Note first that $\frac{\mathbf{E}Z_{2}(n+1)}{\mathbf{E}Z_{2}(n)}=m.$ Hence
we can consider $\widehat{m}_{n}=\frac{Z_{2}(n+1)}{Z_{2}(n)}$ as an
estimator of the parameter $m$ (similar to Lotka-Nagaev estimator for the
classical BGW branching process). It is possible to use also the following
Harris type estimator

$\widetilde{m}_{n}=\sum_{i=2}^{n+1}Z_{2}(i)/%
\sum_{j==1}^{n}Z_{2}(j),n=1,2,....,$ \newline
or Crump and Hove type estimators

$\overline{m}_{n,N}=\sum_{i=n+1}^{n+N}Z_{2}(i)/%
\sum_{j==n}^{n+N-1}Z_{2}(j),n=1,2,...;N=1,2,....$

See $[7]$ for more details.

Estimating $m$ we are able to predict the mean value of the contaminated
(non observed) individuals in the population. In the case when we assume
that $Z_{1}(0)=1$ then $M_{1}(n)=\mathbf{E}Z_{1}(n)$ can be approximated
respectively by\ $\widehat{m}_{n}^{n}$, or $\widetilde{m}_{n}^{n}$, or $%
\overline{m}_{n,N}^{n}.$ In fact it means that we can obtain three types of
estimators

$\widehat{M}_{1}(n)=$\ $\widehat{m}_{n}^{n},$ $\widetilde{M}_{1}(n)=$\ $%
\widetilde{m}_{n}^{n}$ and $\overline{M}_{1}(n)=\overline{m}_{n,N}^{n}.$

In other words we could say that we have at least three prognostic lines.
Therefore if we have observations $(Z_{2}(1),Z_{2}(2),...,Z_{2}(n))$ over
the first $n$ days, we are able to predict the mean value of the
contaminated individuals for the next $k$ days by the relations: \newline
$\widehat{M}_{1}(n+k)=$\ $\widehat{m}_{n}^{n+k},$ $\widetilde{M}_{1}(n+k)=$\ 
$\widetilde{m}_{n}^{n+k}$ and $\overline{M}_{1}(n+k)=\overline{m}%
_{n,N}^{n+k},$ $k=1,2,...$

We are able to estimate also the proportion $\alpha (n)$ of the registered
contaminated individuals among the population in the $n$-th day. Then we can
obtain the following three types of estimators: $\widehat{\alpha }(n)=$ $%
Z_{2}(n)/\{Z_{2}(n)+\widehat{M}_{1}(n)\},$ $\widetilde{\alpha }%
(n)=Z_{2}(n)/\{Z_{2}(n)+\widetilde{M}_{1}(n)\},$ $\overline{\alpha }%
(n)=Z_{2}(n)/\{Z_{2}(n)+\overline{M}_{1}(n)\}.$

All obtained estimators will be presented by the observed registered
lab-confirmed cases in the next section. The quality of the estimation,
however, depends on the representativeness of the sample due to the
specifics of the data collection in each country.

\textbf{Comment. }For more detailed investigation and simulation the
following models can be applied in the considered situation:

$(\mathbf{i)}$\textbf{\ }$h^{\ast
}(s)=q+p_{0}+p_{1}s+p_{2}s^{2}+...+p_{k}s^{k},$ where $q=1-%
\sum_{j=0}^{k}p_{j}$ and $p_{j},j=0,1,...,k,$ can be specially chosen for $%
k=2,3,4,5,6,7,8.$

$(\mathbf{ii)}$ $h^{\ast }(s)=q+p_{0}+\sum_{k=1}^{\infty
}(1-p)p^{k}s^{k}=q+p_{0}+(1-p)ps)/(1-ps),$ where $q+p_{0}=1-p.$ It is
possible to consider also the restricted geometrical distribution up to some 
$k=2,3,4,5,6,7,8.$

$\mathbf{(iii)}$ $h^{\ast }(s)=q+p_{0}+\sum_{k=1}^{\infty }e^{-\lambda }%
\frac{\lambda ^{k}}{k!}s^{k}=q+p_{0}+e^{-\lambda (1-s)}-e^{-\lambda },$
where $q+p_{0}=e^{-\lambda }.$ Similarly it is possible to consider also the
restricted Poisson distribution up to some $k=2,3,4,5,6,7,8.$

Note that the parameters of these distributions can be set in the manner
that $\frac{d}{ds}h^{\ast }(s)|_{s=1}$ is equal to \ $\widehat{m}_{n}^{n}$,
or $\widetilde{m}_{n}^{n}$, or $\overline{m}_{n,N}^{n}.$ Then with this
individual distributions it is possible to simulate the trajectories of the
non-observed process of contamination for further studies.

\textbf{3. Estimation and prediction.}

We would like to point out once again that the considered in Section 2 model
is versitile but the application in each country is specific because it
depends essentially on the official data from the country. The plots and
tables below illustrate well some specific details for different countries
as well as the common trend.

The data used for the estimation of the parameters of the model come from
the official World Health Organization (WHO) situation reports $[10]$.

We apply the here defined model. Note that the observed data is the number
of the newly (daily) registered individuals denoted by $Z_{2}(n)$. The data
about the new number of infected individuals (denoted by $Z_{1}(n)$) is
unobservable. The initial number $m_{0}=EZ_{1}(0)$ is also unknown. Here $n$
is the corresponding day from the beginning of the contamination.

The estimation can be summarized in the following steps.

\begin{enumerate}
\item On the basis of each sample $Z_2(1),\dots,Z_2(s)$, $s = 1,\dots,n$,
the mean numbers of the new infected individuals by one contaminated
individual is estimated by the Harris, Lotka-Nagaev and Crump-Hove type
estimators considered above.

The dynamics of these parameters over time is studied and for the
forecasting purposes the most recent value is used.

\item The mean values of the expected number of nonregistered contaminated
individuals are calculated for the three types of estimators.

Note that $M_1(s+k) = m_0 m^{s+k} = M_1(s) m^k$. Instead of $m_0$ we
estimate $M_1(s)$ by the registered contaminated individuals in day $s$.
Here $s$ depends on the data set and $k=n-s,\dots,n$.

\item The proportion $\alpha (s+k)$ of the registered contaminated
individuals among the population of all infected in the $s+k$-th day is
estimated.
\end{enumerate}

We will demonstrate the approach described above with the data of the
reported laboratory-confirmed COVID-19 daily cases globally, in Italy and in
Bulgaria.

\begin{figure}[]
\begin{tabular}{cc}
\includegraphics[scale = 0.13]{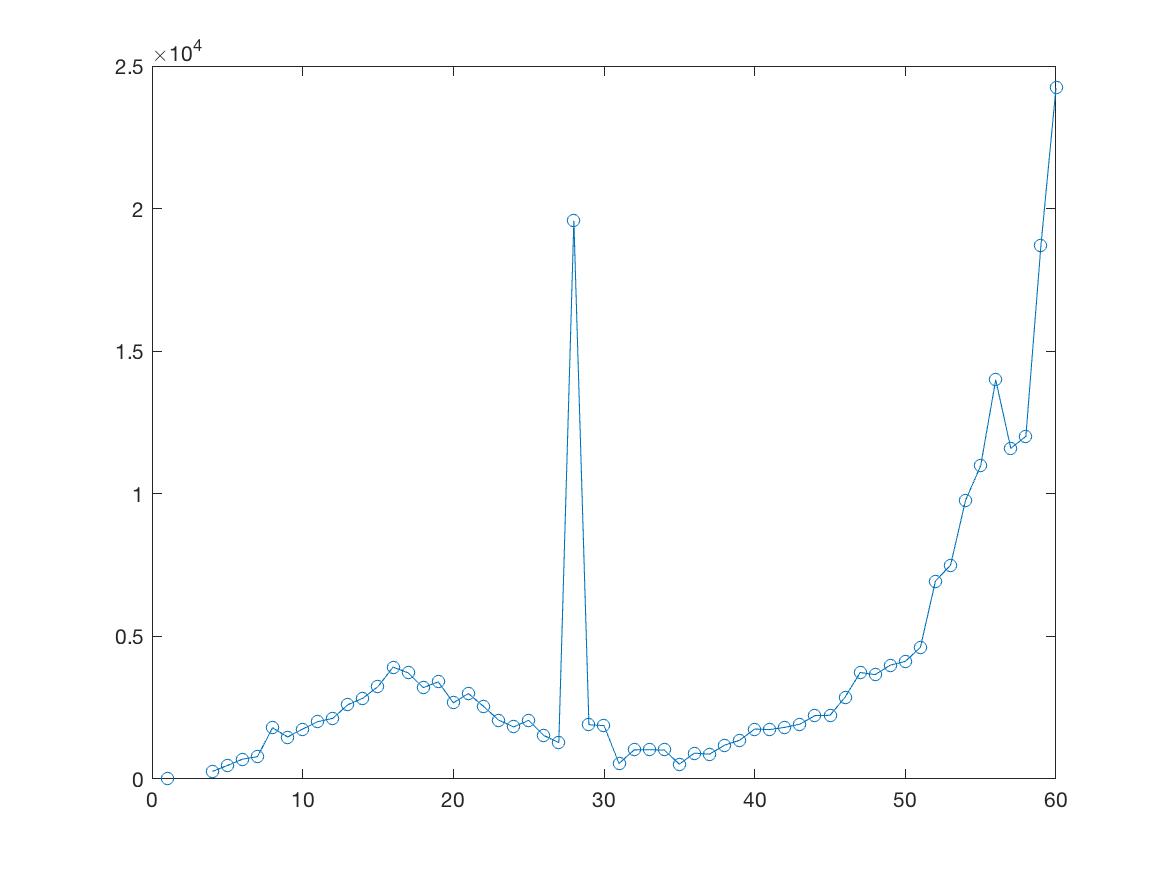} & 
\includegraphics[scale =
0.13]{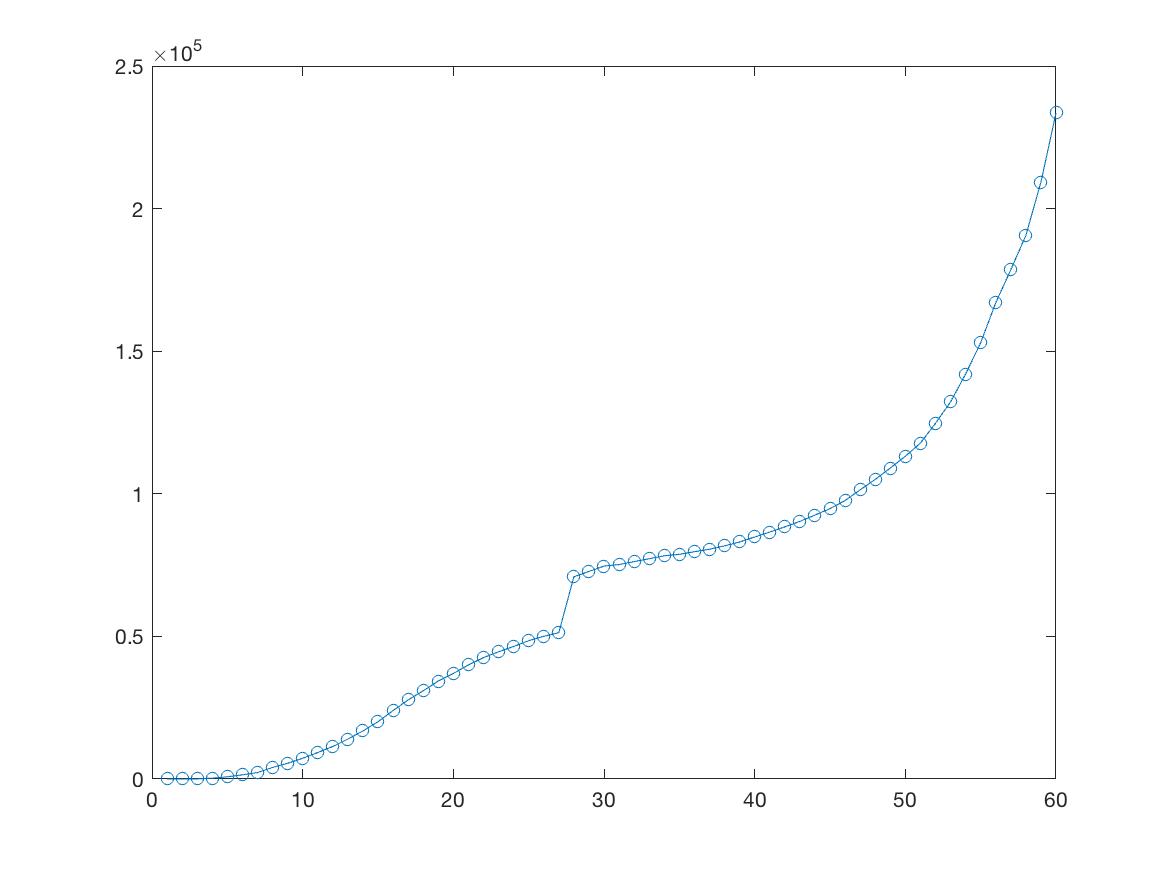} \\ 
{\small New registered} & {\small Total} \\ 
& 
\end{tabular}%
\caption{Raw data for infected globally}
\label{fig:Z2T}
\end{figure}

On Figure \ref{fig:Z2T} one can see the original data for the newly reported
and for the total number of registered cases globally.

\begin{center}
\begin{figure}[]
\begin{center}
\includegraphics[scale = 0.2]{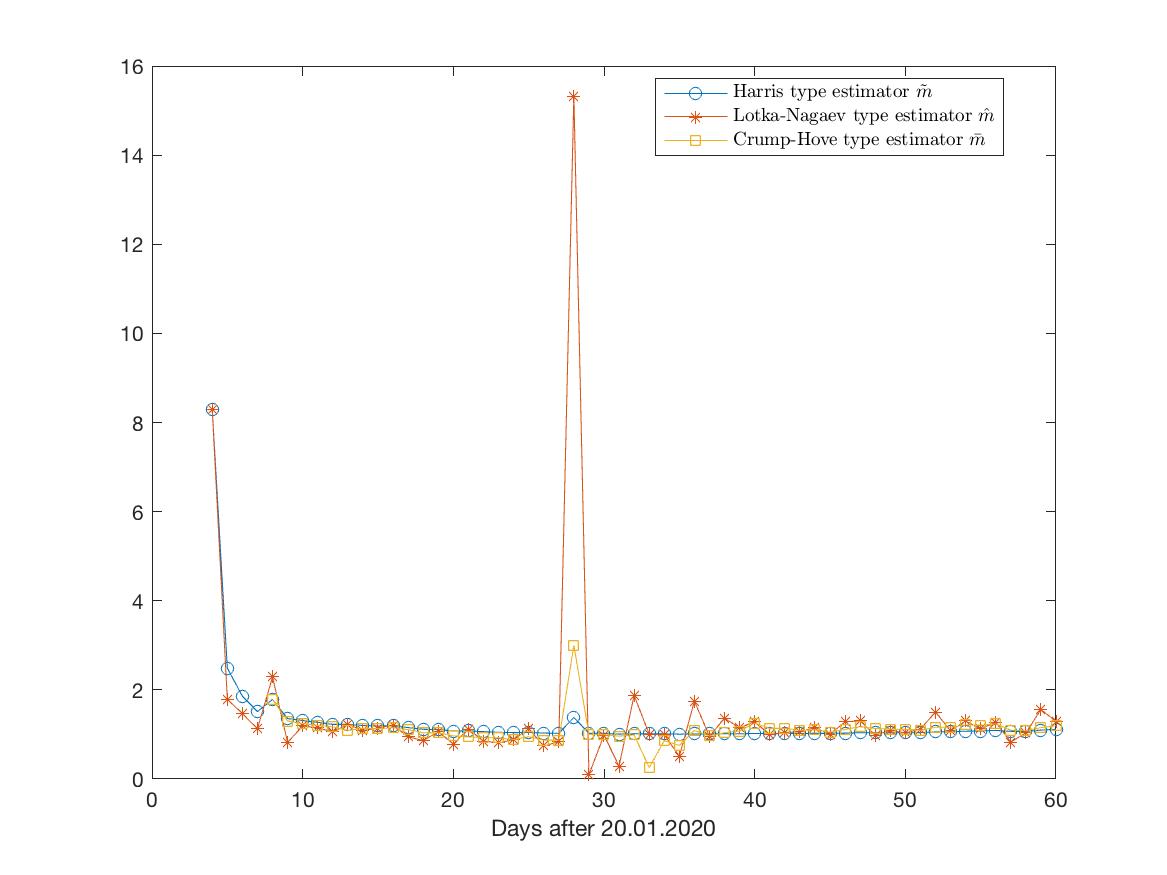}
\end{center}
\caption{{\protect\small Dynamics of the mean numbers of of the new
contaminated individuals}}
\label{fig:mT}
\end{figure}
\end{center}

The dynamics of the Harris, Lotka-Nagaev and Crump-Hove estimators can be
followed on Figure \ref{fig:mT}.

The Harris estimator shows a relatively more stable behaviour during the
period. The three estimates exhibit similar asymptotics.

\begin{center}
\begin{figure}[]
\begin{center}
\includegraphics[scale = 0.2]{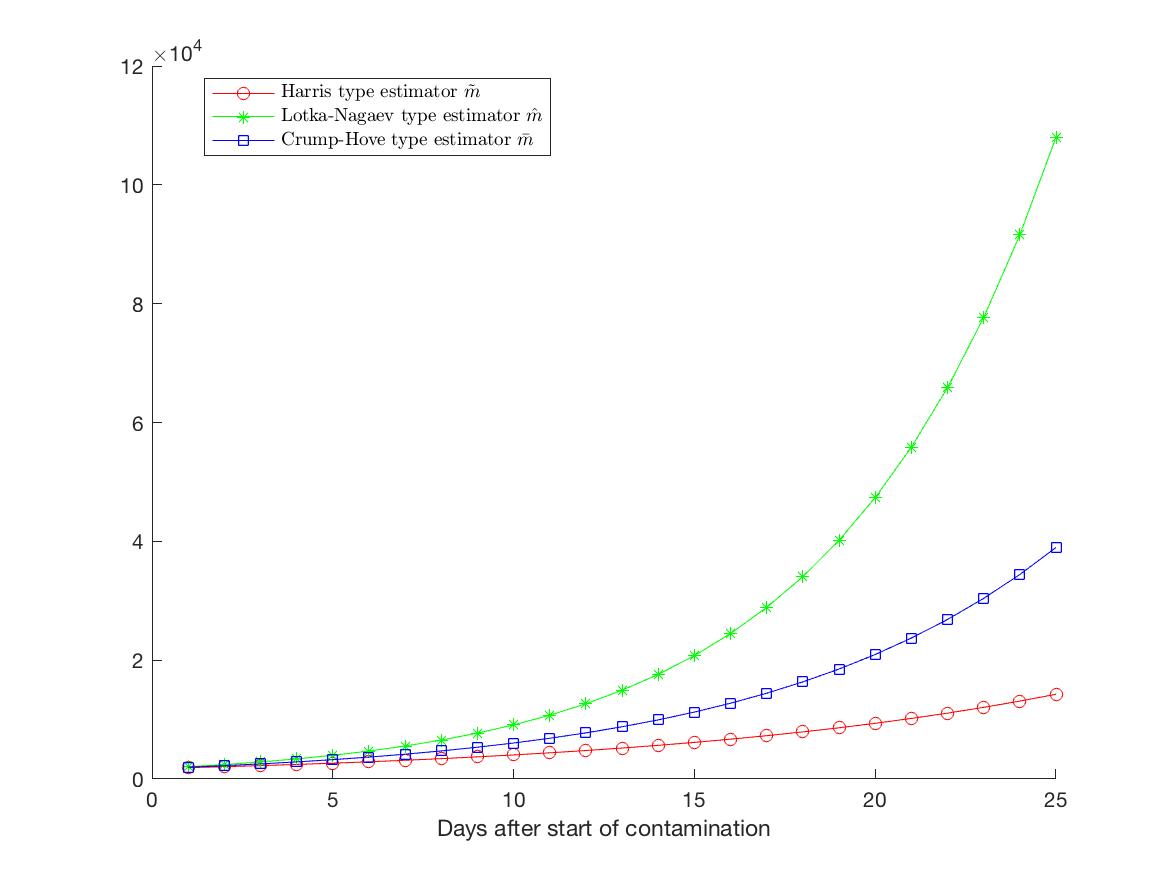}
\end{center}
\caption{{\protect\small The mean values of the expected number of
nonregistered}}
\label{fig:M1T}
\end{figure}
\end{center}

The next step is to calculate the mean values of the expected number of
nonregistered contaminated individuals for the three types of estimators,
starting with $s=20$ (Figure \ref{fig:M1T}). The last five points on the
graph after day 20 represent the forecast for the next 5 days.

\begin{center}
\begin{figure}[]
\begin{center}
\includegraphics[scale = 0.2]{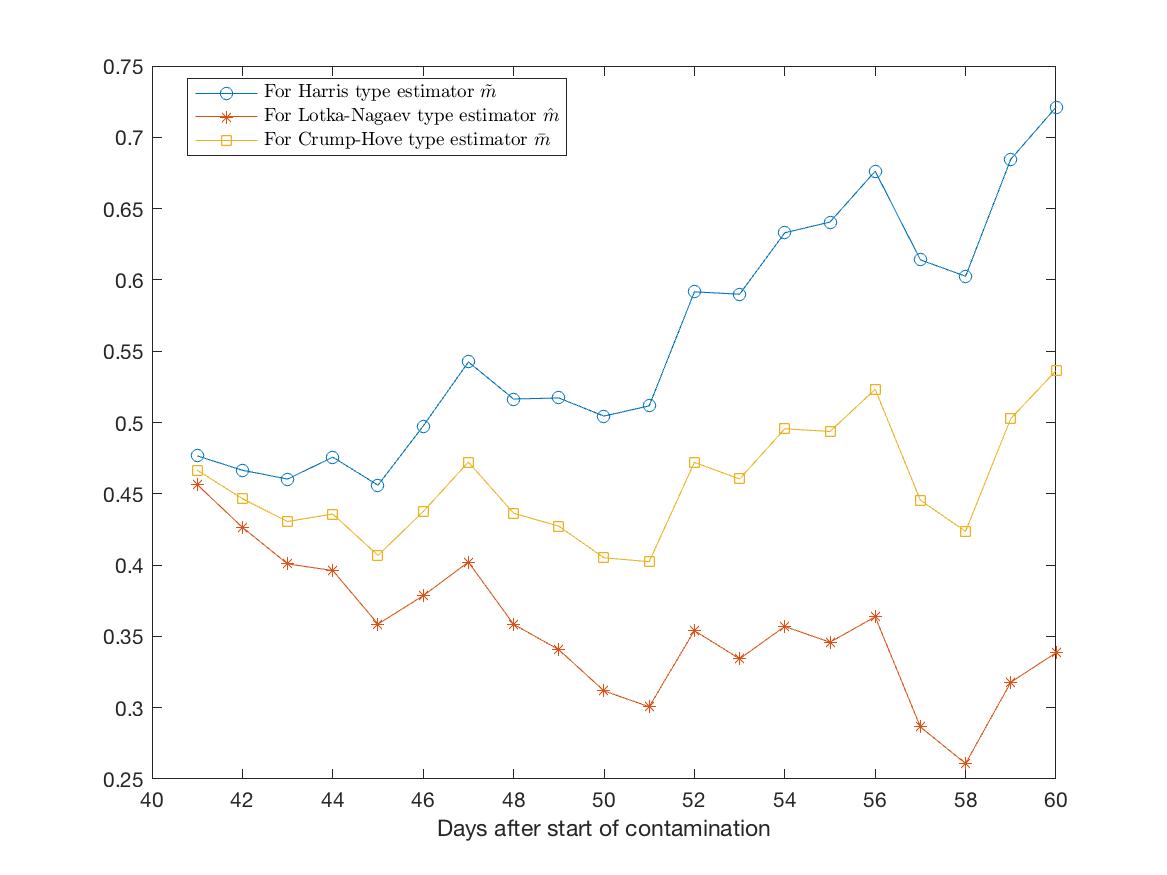}
\end{center}
\caption{{\protect\small The proportion of the registered individuals}}
\label{fig:aT}
\end{figure}
\end{center}

The corresponding values of the proportion of the registered individuals
among the contaminated population is presented on Figure \ref{fig:aT}

For the registered cases in Italy the raw data can be seen on the Figures %
\ref{fig:Z2I} - \ref{fig:aI} below.

\begin{figure}[]
\begin{tabular}{cc}
\includegraphics[scale = 0.13]{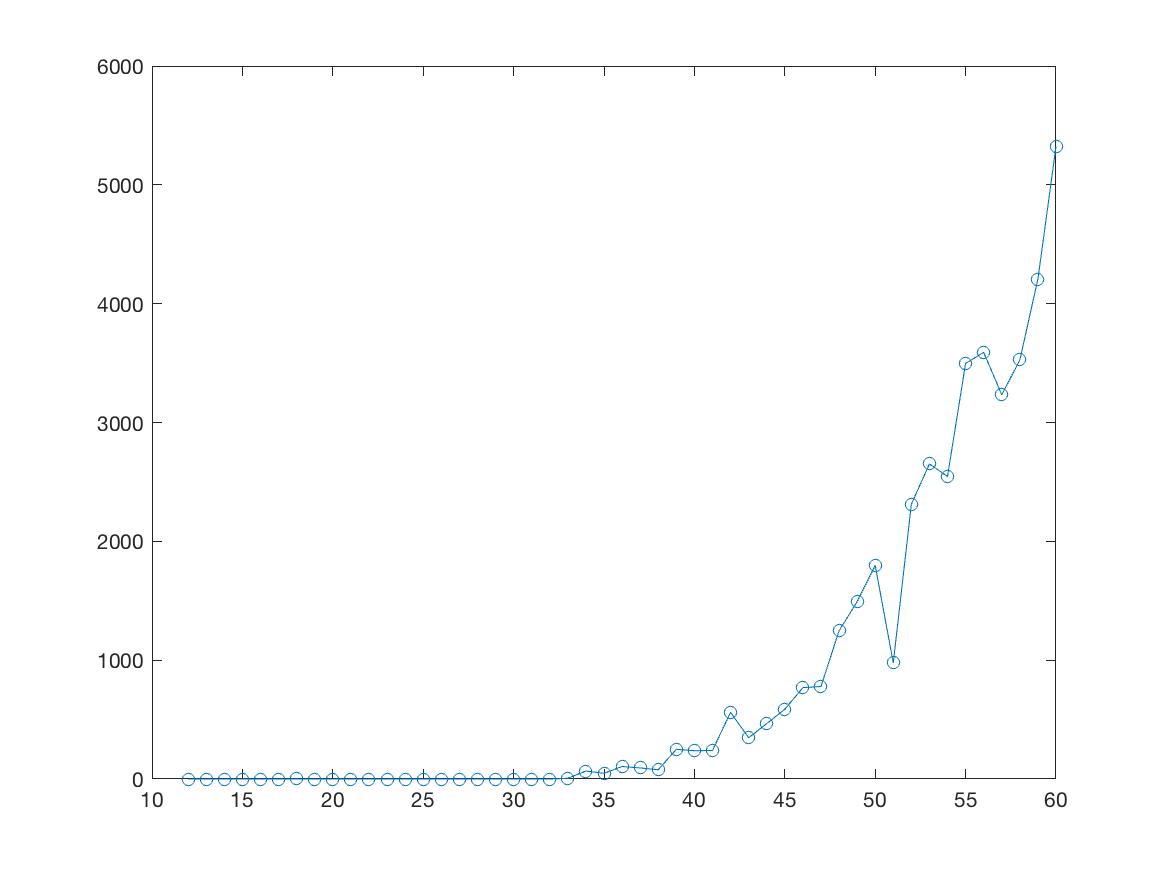} & 
\includegraphics[scale =
0.13]{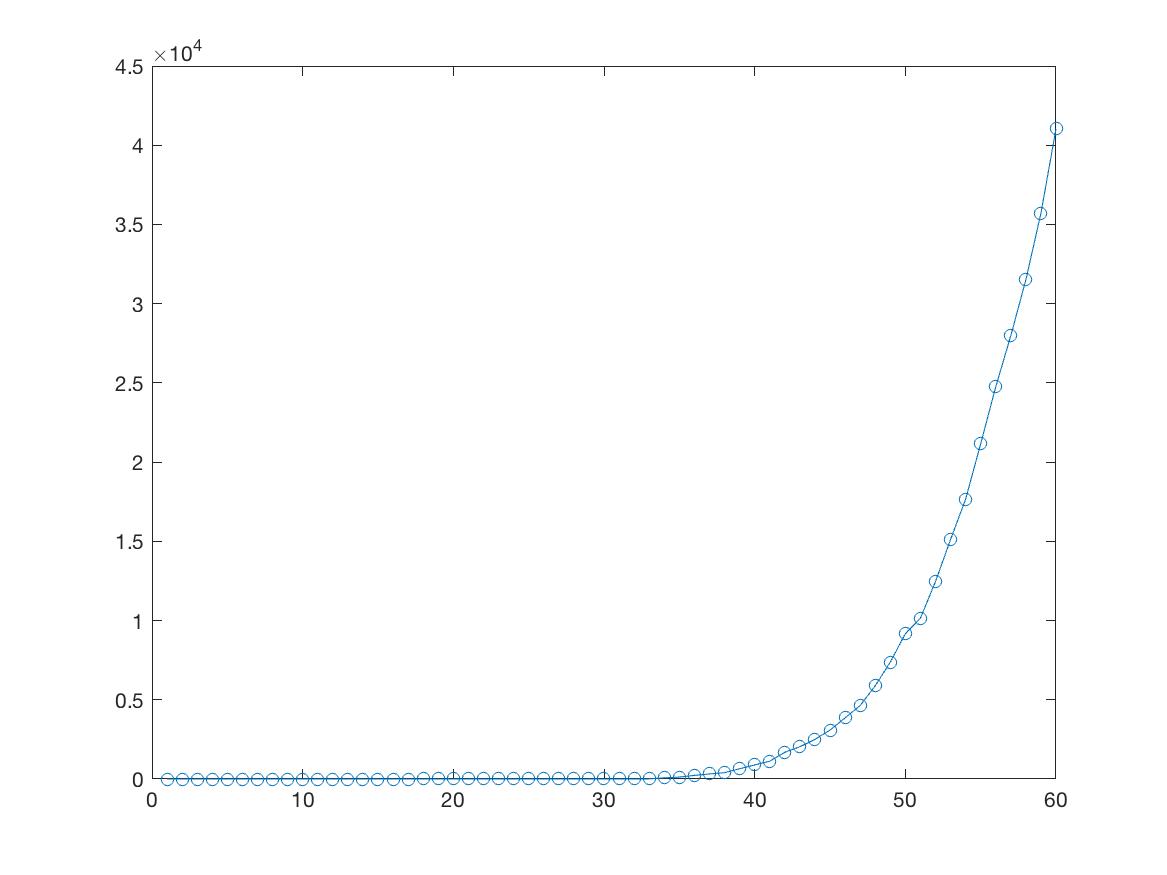} \\ 
{\small New registered} & {\small Total} \\ 
& 
\end{tabular}%
\caption{Raw data, infected, Italy}
\label{fig:Z2I}
\end{figure}

\begin{center}
\begin{figure}[]
\begin{center}
\includegraphics[scale = 0.2]{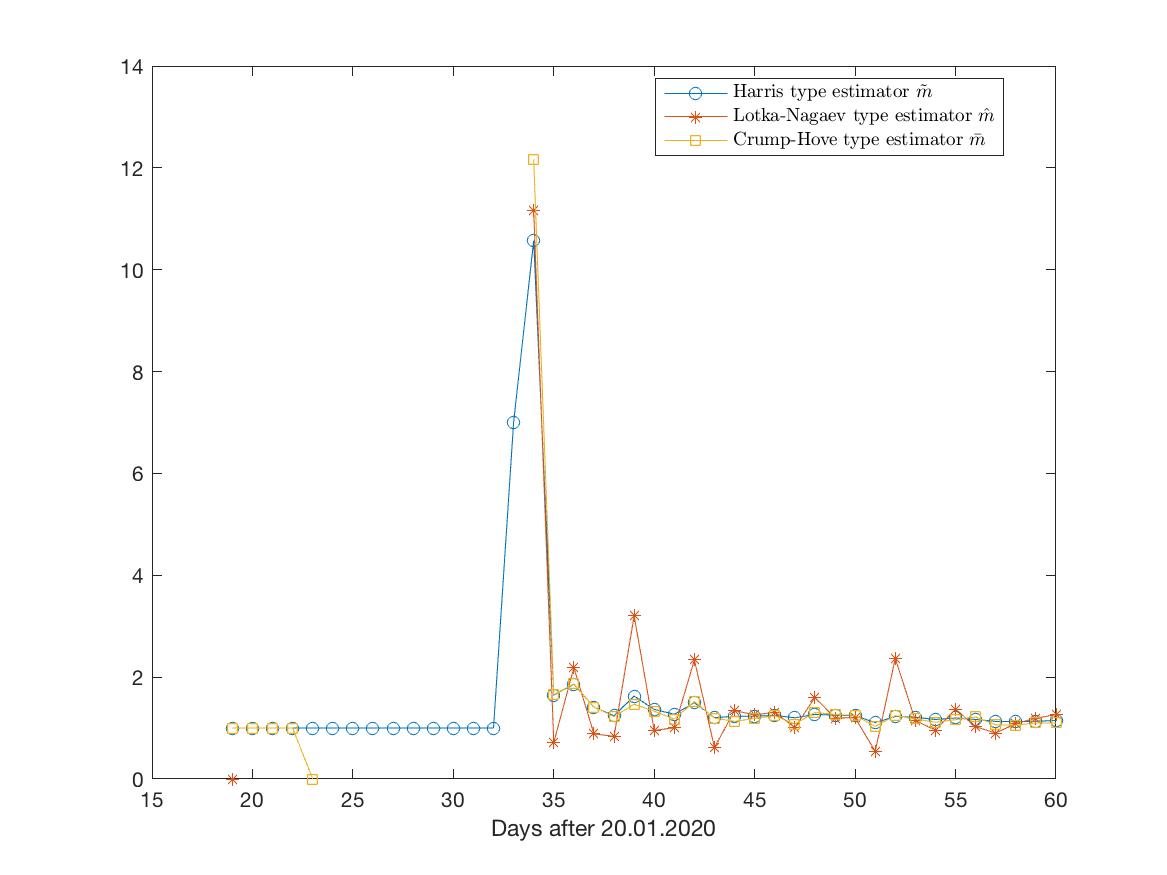}
\end{center}
\caption{{\protect\small Dynamics of the mean numbers of of the new
contaminated individuals}}
\label{fig:mI}
\end{figure}

\begin{figure}[]
\begin{center}
\includegraphics[scale = 0.2]{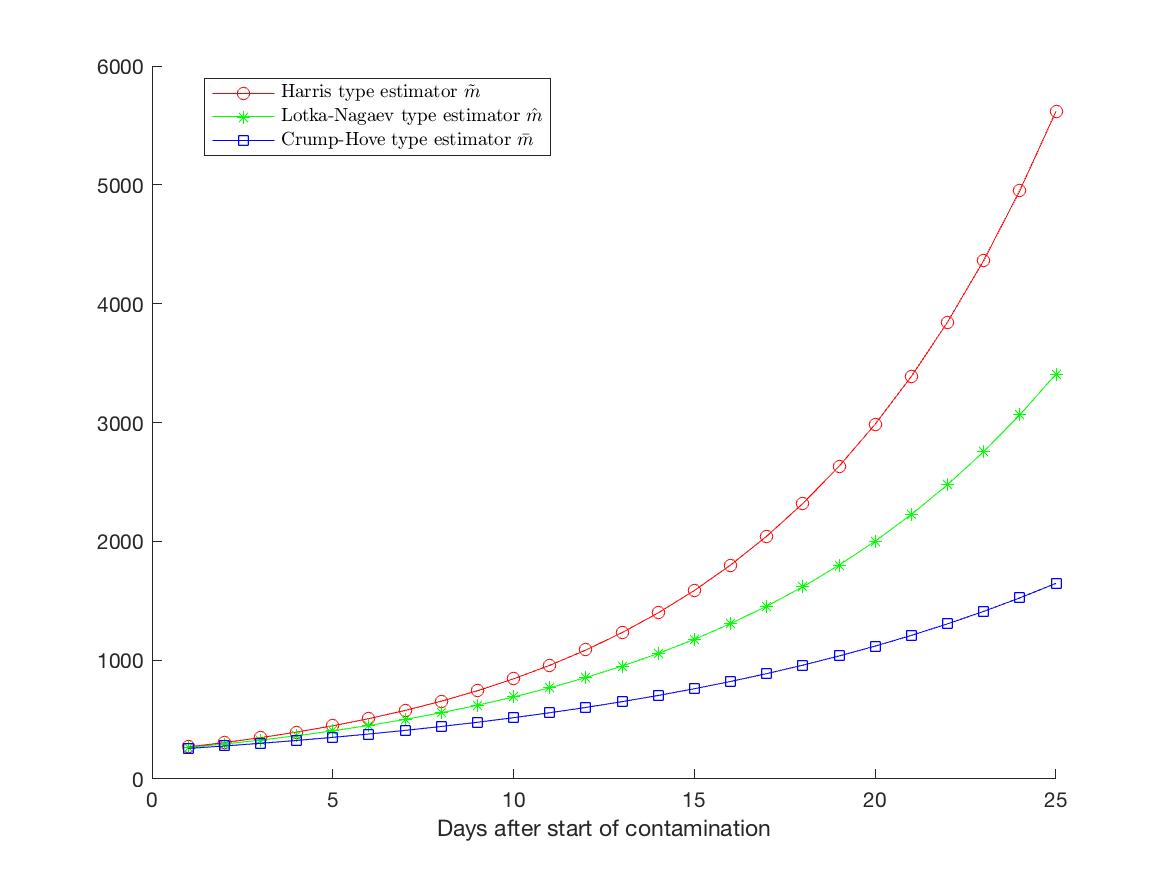}
\end{center}
\caption{{\protect\small The mean values of the expected number of
nonregistered}}
\label{fig:M1I}
\end{figure}

\begin{figure}[]
\begin{center}
\includegraphics[scale = 0.2]{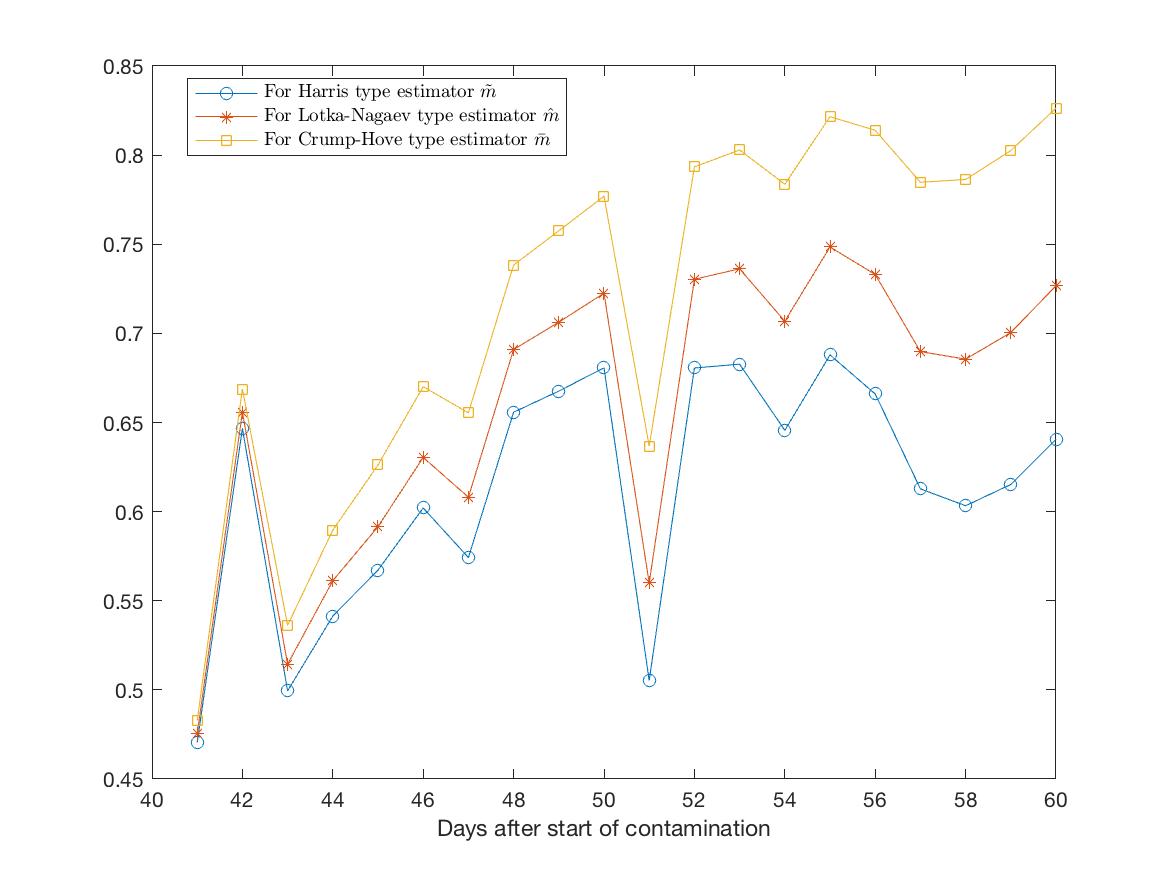}
\end{center}
\caption{{\protect\small The proportion of the registered individuals}}
\label{fig:aI}
\end{figure}
\end{center}

Note that the newly confirmed contaminated individuals in Bulgaria are
reported day by day, starting from 08.03.2020 as:\newline
\begin{tabular}{c}
$4;0;2;1;16;8;10;10;11;19;11;18;17;36;22;16;19;22;22;29;38$%
\end{tabular}

In fact these data form the sample $(Z_{2}(1),Z_{2}(2),...,Z_{2}(20)), $
where now $n=20,$ i.e. $20$ days from the first confirmed contaminated
individuals. We use also the statistics $U(k)=%
\sum_{j=1}^{k}Z_{2}(j),k=1,2,...,17,$ whose values are given as follows:%
\newline
\begin{tabular}{c}
$4;4;6;7;23;31;41;51;62;81;92;110;127;163;185;201;220;242;264;293;331.$%
\end{tabular}

\begin{figure}[]
\begin{tabular}{cc}
\includegraphics[scale = 0.13]{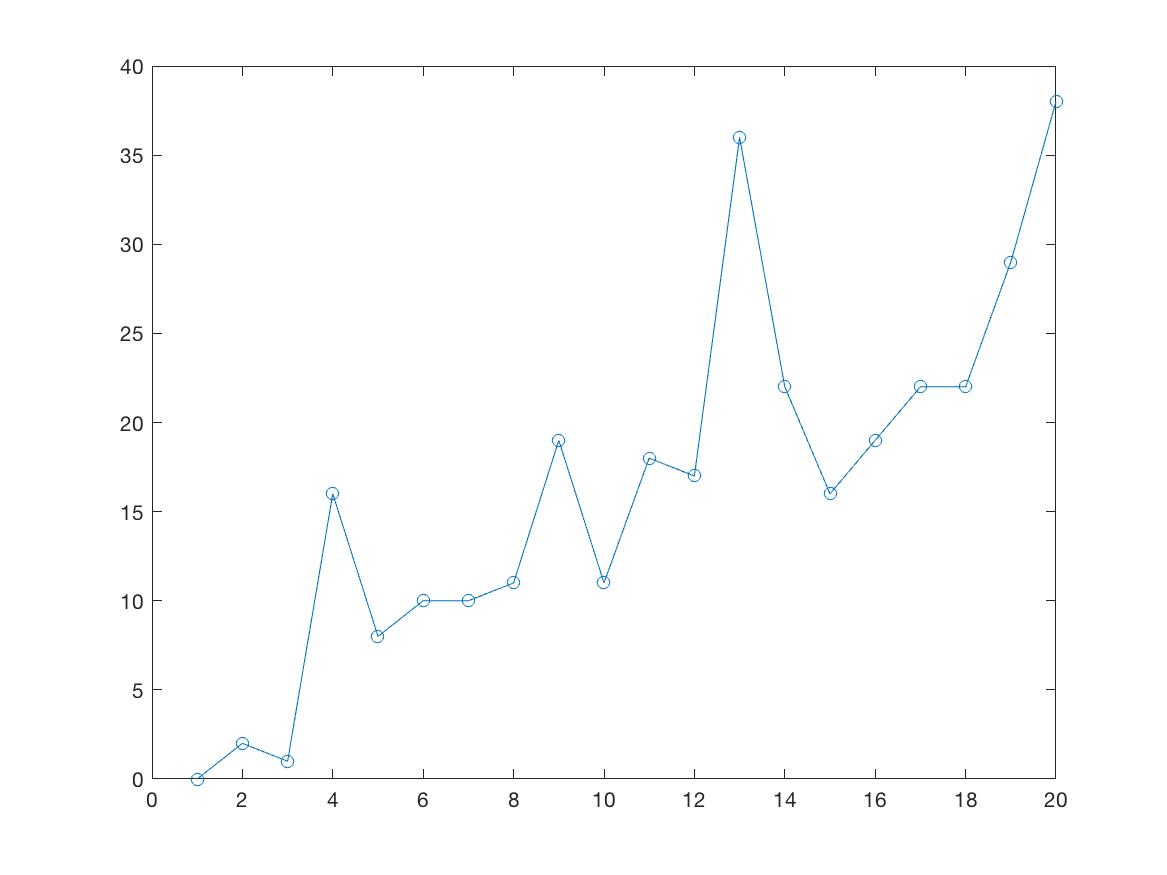} & 
\includegraphics[scale =
0.13]{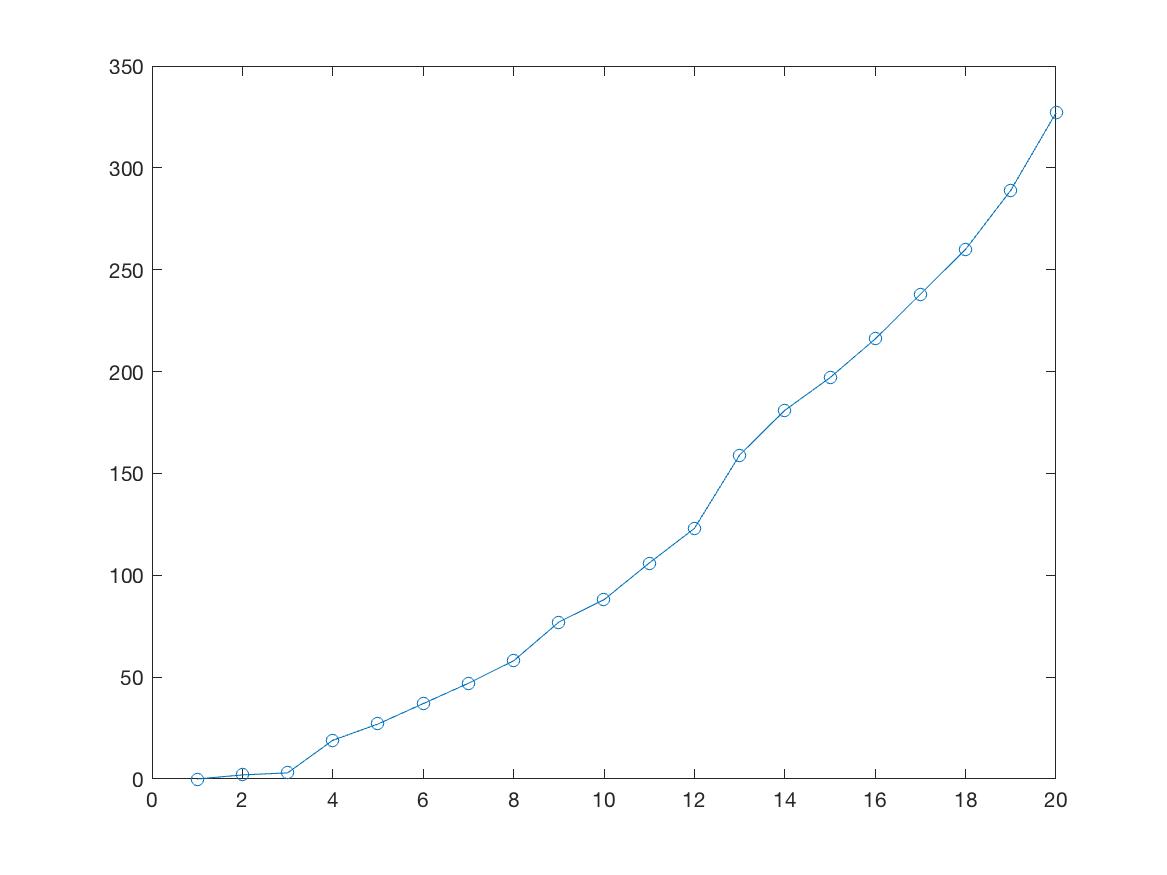} \\ 
{\small New registered} & {\small Total} \\ 
& 
\end{tabular}%
\caption{Raw data, infected, Bulgaria}
\label{fig:Z2B}
\end{figure}

The two samples are presented on Figure \ref{fig:Z2B}.

The estimated mean number of infected individuals for the three types of
estimators is presented on Figure \ref{fig:mB}, the estimated expected
number of contaminated individuals (with $s=10$, chosen on the basis of the
shorter time period of contamination) can be seen on Figure \ref{fig:M1B}
and the estimated $\alpha$ - on Figure \ref{fig:aB}.

\begin{center}
\begin{figure}[]
\begin{center}
\includegraphics[scale = 0.2]{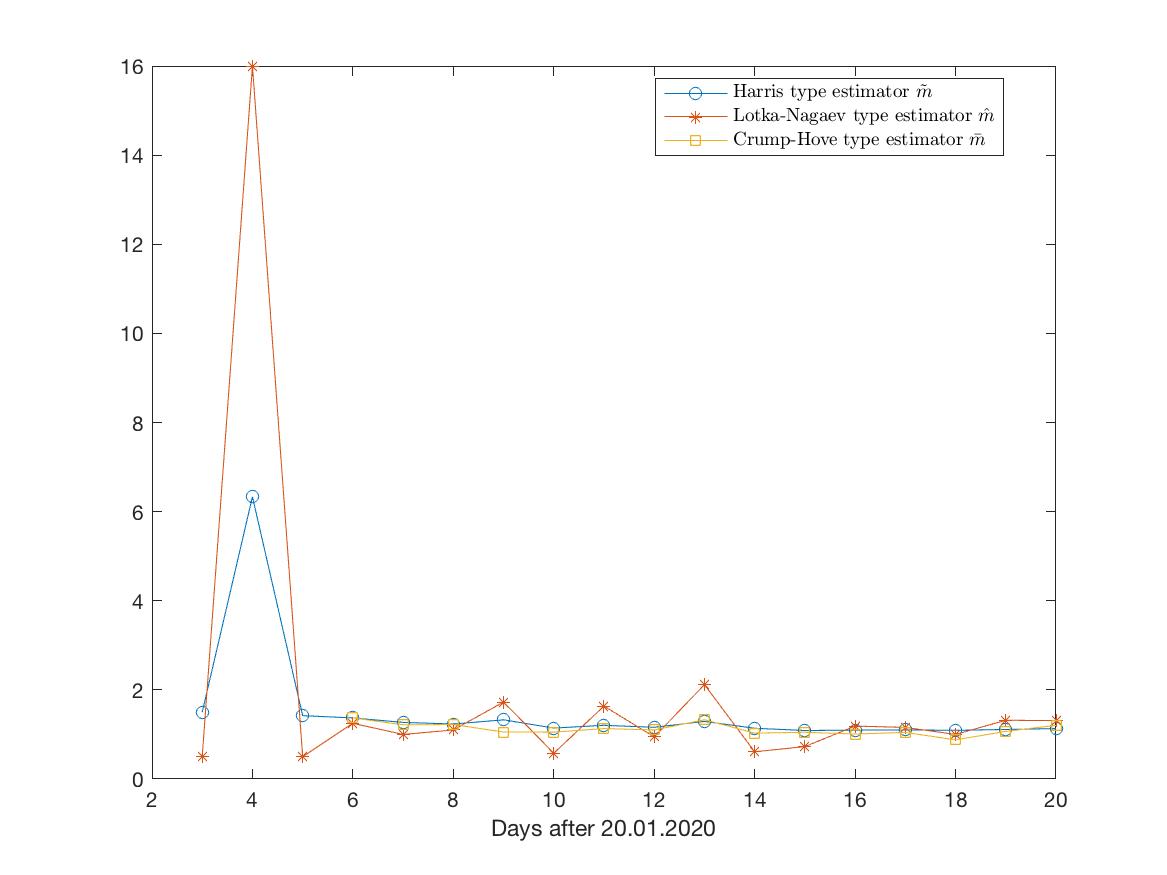}
\end{center}
\caption{{\protect\small Dynamics of the mean numbers of of the new
contaminated individuals}}
\label{fig:mB}
\end{figure}

\begin{figure}[]
\begin{center}
\includegraphics[scale = 0.2]{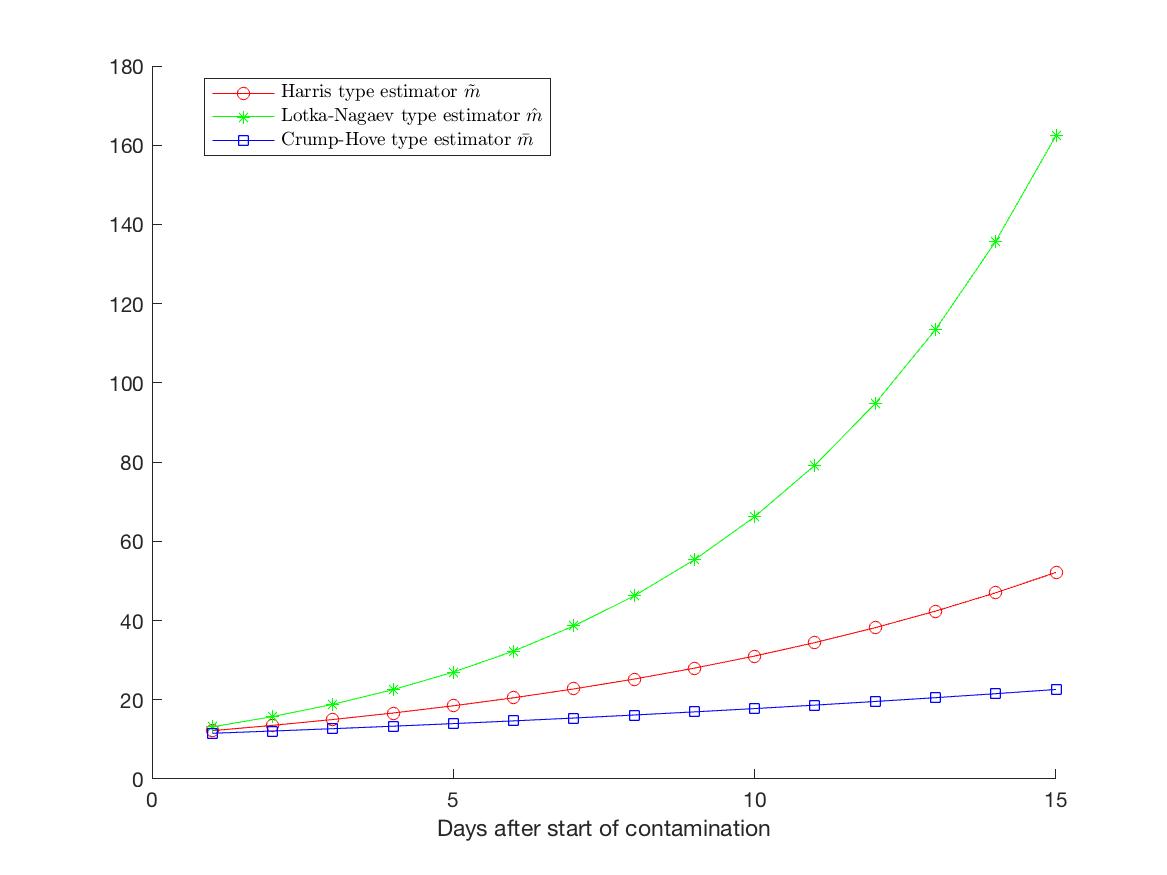}
\end{center}
\caption{{\protect\small The mean values of the expected number of
nonregistered}}
\label{fig:M1B}
\end{figure}

\begin{figure}[]
\begin{center}
\includegraphics[scale = 0.2]{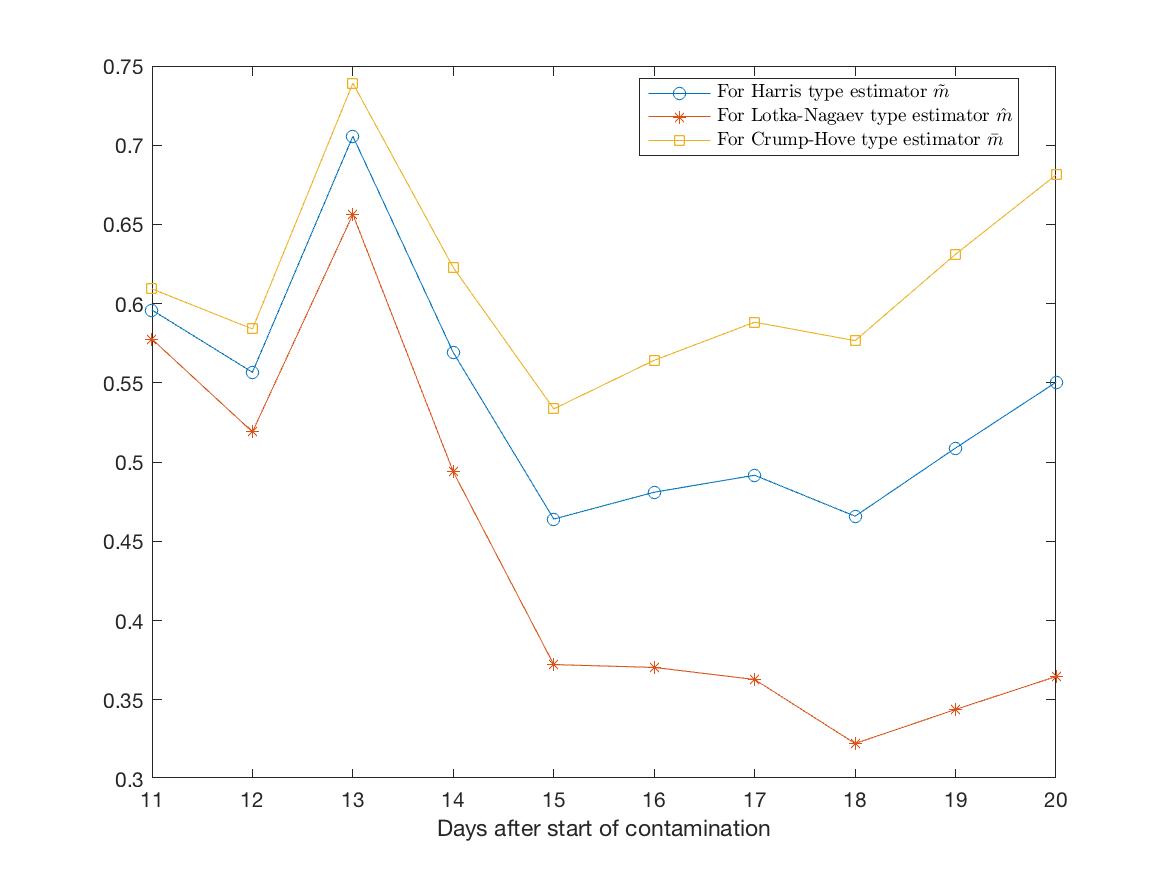}
\end{center}
\caption{{\protect\small The proportion of the registered individuals}}
\label{fig:aB}
\end{figure}
\end{center}

The values of the Harris estimator at the end of the contamination period
for the three datasets considered above can be found in Table \ref{tab:m}

\begin{table}[]
\begin{center}
\begin{tabular}{|c|c|c|c|}
\hline
\textbf{Data set} & \textbf{$\tilde m_n$} & \textbf{95\% CI lower} & \textbf{%
95\% CI upper} \\ \hline
Globally & 1.0875 & 0.9344 & 1.2406 \\ 
Italy & 1.1348 & 1.0694 & 1.2002 \\ 
Bulgaria & 1.1093 & 0.8845 & 1.3341 \\ \hline
\end{tabular}%
\end{center}
\caption{Values of the Harris estimator}
\label{tab:m}
\end{table}

Up to day 20 the Harris estimator (for Bulgaria) is 1.1093 (i.e. slightly
supercritical), the expected number of newly infected and still not
registered individuals is appr. 31, the actual reported registered is 38,
hence the estimated proportion of the registered individuals among the
infected is $\alpha =0.5507$.

To approve the behaviour of the model one can compare the predicted number
of confirmed individuals $\tilde Z_2(n-k)$ (using the Harris estimator $%
\tilde m_n$) for the period of $k$ days before the latest date in the data
set with the actually observed values of $Z_2(n)$. The corresponsing values,
as well as the 95\% confidence interval of the forecast are presented on
Table \ref{tab:forecast}.

\begin{table}[]
\begin{center}
\begin{tabular}{|c|c|c|c|c|c|}
\hline
\textbf{Data set} & {day $k$} & \textbf{$\tilde Z_2(n-k)$} & \textbf{$%
Z_2(n-k)$} & \textbf{95\% CI lower} & \textbf{95\% CI upper} \\ \hline
Globally & 5 & 11754 & 13998 & 8289 & 15393 \\ 
& 4 & 15128 & 11596 & 10364 & 19757 \\ 
& 3 & 12515 & 12016 & 8454 & 16575 \\ 
& 2 & 12928 & 18712 & 8556 & 17416 \\ 
& 1 & 20174 & 24234 & 12802 & 27250 \\ \hline
Italy & 5 & 4172 & 3590 & 3460 & 4444 \\ 
& 4 & 4232 & 3233 & 3555 & 4646 \\ 
& 3 & 3770 & 3526 & 3236 & 4304 \\ 
& 2 & 4027 & 4207 & 3532 & 4780 \\ 
& 1 & 4754 & 5322 & 4196 & 5843 \\ \hline
Bulgaria & 5 & 19 & 19 & 10 & 25 \\ 
& 4 & 21 & 22 & 11 & 31 \\ 
& 3 & 24 & 22 & 12 & 37 \\ 
& 2 & 24 & 29 & 11 & 38 \\ 
& 1 & 32 & 38 & 15 & 53 \\ \hline
\end{tabular}%
\end{center}
\caption{Observed and predicted registered cases}
\label{tab:forecast}
\end{table}

Additionaly, using the data, provided by WHO the comparison between the
behaviour of the process under different local conditions with a
approximatelly same contamination period. For example on the Figure \ref%
{fig:mCompare} the Harris estimator globally, for Italy, Germany and France
are compared.

As mentioned earlier the mean number of new contaminated individuals by one
c.i. tends to values close, but greater than 1 as the time of contamination
increases.

Calculating the proportions $\alpha$ of the confirmed contaminated
individuals among the four populations with $s=20$ (presented on Figure \ref%
{fig:aCompare}) one can see that $\alpha$'s vary largely but stay relatively
high, especially at the end of the contamination period. This fact can be
considered as a result of different types of actions, undertaken to restrict
the spread of the infection.

\begin{center}
\begin{figure}[]
\begin{center}
\includegraphics[scale = 0.2]{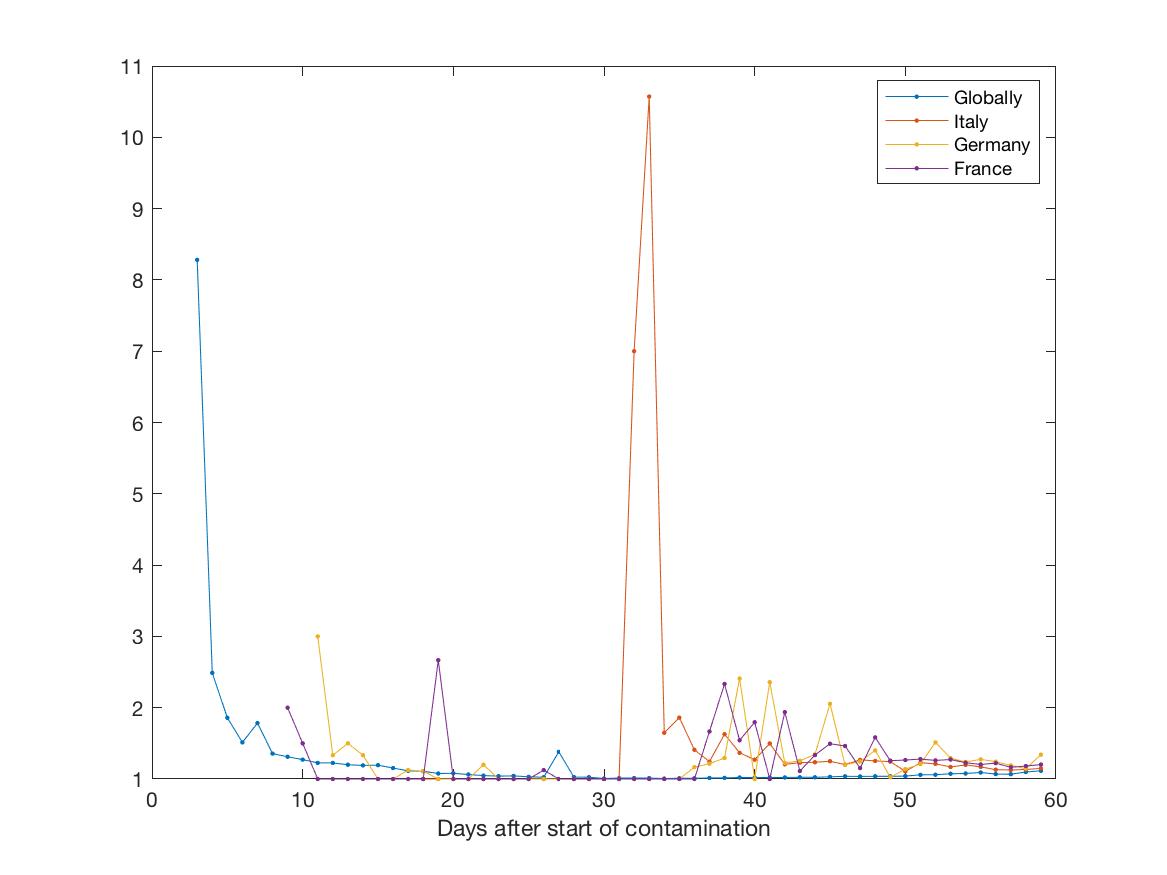}
\end{center}
\caption{{\protect\small Harris estimator}}
\label{fig:mCompare}
\end{figure}

\begin{figure}[]
\begin{center}
\includegraphics[scale = 0.2]{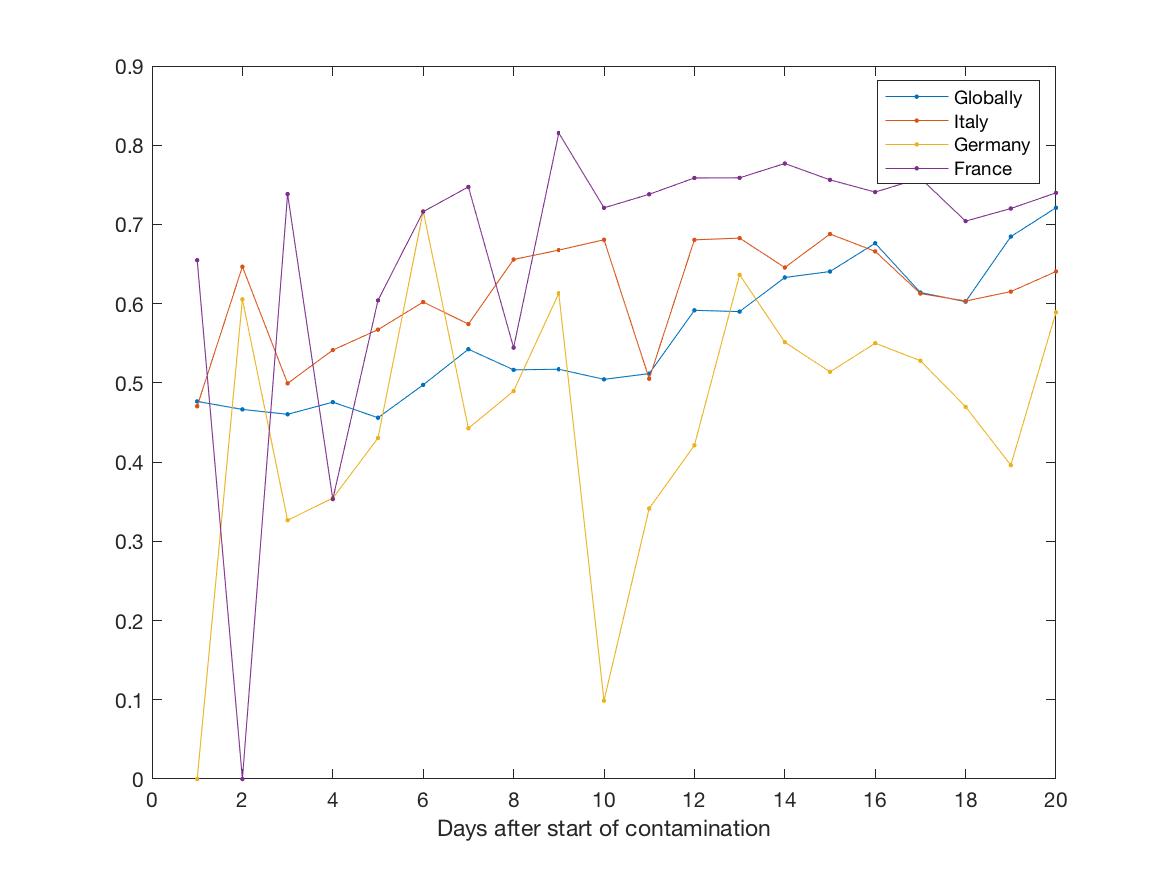}
\end{center}
\caption{{\protect\small The proportion of the registered individuals in the
last 20 days}}
\label{fig:aCompare}
\end{figure}
\end{center}

\textbf{4. Concluding remarks.}

First of all the estimation of the mean value of reproduction $m$ allows us
to classify the contamination process as supercritical ($m>1$), critical ($%
m=1$) and subcritical ($m<1$). In the supercritical case the mean population
growth is exponential, in the critical case the mean value of the population
is constant and in the subcritical case the decreasing of the mean
population is exponential.

Up to the moment of our investigation the estimated mean number of new
contaminated individuals (for Bulgaria) is slightly greater than 1 which
corresponds to the exponential growth of the contaminated population,
globally and locally in specific countries and regions.

Finally the estimating of the mean parameter of contamination can be
considered as a first stage to construction of a more complicated
epidemiological model. As a such model for example, one can use a branching
process with random migration considered in [8-9] or some other model of
controlled branching processes (see [5]).

Additional information, reports and plots, related to this research can be
found on http://ir-statistics.net/covid-19.

\textbf{References}

1. Harris, T.E. The Theory of Branching Processes. Springer, Berlin, 1963.

2. Sevastyanov, B.A. Branching Processes. Nauka, Moscow, 1971. (In Russian).

3. Athreya, K.B., P.E. Ney. Branching Processes. Springer, Berlin, 1972.

4. Jagers, P. Branching Processes with Biological Applications. Wiley,
London,1975.

5. Gonzalez, M., I.M. del Puerto, G.P. Yanev. Controlled Branching
Processes. Wiley, London, 2018.

6. Yakovlev, A. Yu., N. M. Yanev. \textit{Transient Processes in Cell
Proliferation Kinetics.} Lecture Notes in Biomathematics 82, Springer, New
York, 1989.

7. Yanev, N.M. Statistical inference for branching processes, Ch.7 (143-168)
in: Records and Branching processes, Ed. M.Ahsanullah, G.P.Yanev, Nova
Science Publishers, Inc., New York, 2008.

8. Yanev,G.P., N.M. Yanev. Critical branching processes with random
migration. In: C.C. Heyde (Editor), Branching Processes (Proceedings of the
First World Congress). Lecture Notes in Statistics, 99, Springer-Verlag, New
York, 1995, 36-46.

9. Yanev, G.P., N.M. Yanev. Branching Processes with two types of emigration
and state-dependent immigration. In: Lecture Notes in Statistics 114,
Springer-Verlag, New York, 1996, 216-228.

10.
https://www.who.int/emergencies/diseases/novel-coronavirus-2019/situation-reports/ (visited on 20.03.2020)\bigskip

$^{1}$Institute of Mathematics and Informatics, Bulgarian Academy of Sciences

$^{2}$Faculty of Mathematics and Informatics, Sofia University

$^{3}$New Bulgarian University

\end{document}